\documentclass[letter]{ieice}

\usepackage[dvips]{graphicx}
\newenvironment{numberedlist}
{\begin{list}{\makebox[20pt]{\hss(\arabic{itemno})\enspace}}
             {\usecounter{itemno}\labelwidth 20pt}}{\end{list}}

\newcounter{itemno}

\newcounter{itemno1}

\newcounter{itemno2}

\newcounter{exno}

\newcounter{defno}

\newenvironment{defn}{\refstepcounter{defno}\medskip \noindent {\bf
Definition \thedefno.\ }}{\medskip}

\newcommand{\sep}{\;\vert\;}

\newcommand{\oprove}{\vdash\kern-.6em\lower.7ex\hbox{$\scriptstyle O$}\,}

\newcommand{\pderivation}{{\cal P}\kern -.1em\hbox{\rm -derivation}}
\newcommand{\pderivationl}{{\cal P}\kern -.1em\hbox{\em -derivation}}
\newcommand{\pderivable}{{\cal P}\kern -.1em\hbox{\rm -derivable}}
\newcommand{\pderivablel}{{\cal P}\kern -.1em\hbox{\em -derivable}}
\newcommand{\pderivations}{{\cal P}\kern -.1em\hbox{\rm -derivations}}
\newcommand{\pderivability}{{\cal P}\kern -.1em\hbox{\rm -derivability}}

\newcommand{\ie}{{\em i.e.}}

\newsavebox{\lpartfig}
\newsavebox{\rpartfig}

\newenvironment{exmple}{
 \begingroup \begin{tabbing} \hspace{2em}\= \hspace{3em}\= \hspace{3em}\=
\hspace{3em}\= \hspace{3em}\= \hspace{3em}\= \kill}{
 \end{tabbing}\endgroup}

\newcommand{\lb}{\langle}
\newcommand{\rb}{\rangle}

\newcommand{\intp}{intp_o}
  
\newcommand{\add}{\oplus} 
\newcommand{\adc}{\&} 

     {\\* \hspace*{\fill} \end{trivlist}}

\field{}

\title{Extending  Functional Languages with High-level Exception Handling}
\authorlist{
\authorentry{Keehang Kwon}{m}{labelA}
\authorentry{Daeseong Kang}{n}{labelB}
}
\affiliate[labelA]{The author is a professor of  
Computer Eng., DongA University. email:khkwon@dau.ac.kr}
\affiliate[labelB]{The author is a professor of  
Electronics Eng., DongA University. The corresponding author.}

\received{2003}{1}{1}
\revised{2003}{1}{1}
\finalreceived{2003}{1}{1}
\renewcommand{\add}{\bigtriangledown} 
\renewcommand{\adc}{\bigtriangleup} 

\begin{document}
\maketitle
\begin{summary}
We extend functional languages with  new exception handling.
To be specific, we allow 
   sequential-disjunction expressions 
 $E_0 \add \ldots \add E_n$  where  $E_0,\ldots,E_n$ are
expressions.   These expressions have the following intended semantics:
 sequentially $choose$ the first successful  $E_i$ and evaluate $E_i$ where $i$ is among $0,\ldots,n$.  
 These expressions thus allow us to specify an expression $E_i$ with the failure-handling 
(exception handling) 
routine, \ie, expression $E_{i+1}$ for $i = 0,\ldots,n-1$.
We also discuss
sequential-conjunction function declarations $D_0 \adc \ldots \adc D_n$. The latter can be
seen as a dual of sequential-disjunction expressions.

\end{summary}

\newcommand{\muprolog}{{Rec$^{\add,\adc}$}}

\renewcommand{\intp}{eval} 


\section{Introduction}\label{sec:intro}

The theory of recursive functions ($Rec$) provides a basis for functional programming.
It includes operations of composition, recursion, $etc$. Although $Rec$ is quite expressive,
it does not contain error-handling mechanisms.

To fix this problem, we propose to add the following:

\begin{itemize}

\item sequential-disjunction expressions (SD expressions),
originally introduced in the seminal work of Japaridze
\cite{Jap08}, $E_0 \add   \ldots \add E_n$ 
where $E_0,\ldots, E_n$ are expressions. 

\item sequential-conjunction function declarations (SC declarations)
$D_0 \adc   \ldots \adc D_n$ 
  where $D_0,\ldots, D_n$ are function declarations.
\end{itemize}
  
\noindent 
Evaluating  $E_0 \add   \ldots \add E_n$ with respect to a program $D_0 \adc   \ldots \adc D_m$
 --- $\intp(D_0 \adc   \ldots \adc D_m,E_0 \add  \ldots \add E_n)$ ---
 has the following intended semantics: 
 sequentially choose the first successful  one  among
 \[ \intp(D_0 \adc \ldots \adc D_m, E_0),\ldots,\intp(D_m, E_0),\]
 \begin{center}
   $\vdots$
   \end{center}
\[ \intp(D_0 \adc   \ldots \adc D_m, E_n),\ldots, \intp(D_m, E_n). \]
If none of them are successful, just the failure value ($\bot(errcode)$) with its error code
is returned, 
which can be regarded as an uncaught exception.
Thus, SD expressions are intended to deal with exceptions  in the course of evaluating expressions.
On the other hand,  SC declarations -- a dual of SD expressions --
are intended to deal with exceptions occuring in declarations.

    This paper proposes  \muprolog, an extension of the core functional languages with SC/SD  operators. 
    The remainder of this paper is structured as follows. We describe SC expressions and
    SD declarations in the next two sections. We describe \muprolog\
  in Section \ref{s:logic}. In Section \ref{sec:modules}, we
present some examples of  \muprolog.
Section~\ref{sec:conc} concludes the paper.

\section{Sequential-disjunction expressions}

    An
illustration of this  aspect is provided by the following definition of the
function $sort(X)$ where $X$ is a list:

\begin{exmple}
$      sort(X)$ ${\rm =}$ \> \hspace{5em}   \\
  \> $heapsort(X)\ \add\ quicksort(X) \add bubblesort(X)$ 
\end{exmple} 
\noindent
The body of the definition above contains a SD expression, denoted by $\add$.
 As a particular example, evaluating $sort([3,100,40,2])$ would result in selecting and 
executing the first expression $heapsort([3,100,40,2])$.  If the heapsort module is available in the program, then the given goal 
will succeed, producing the solution. If the execution fails for some reason, the machine tries the plan B, \ie, the quicksort module, and
so on.

     As seen from the example above, SD expressions of the form $E_0 \add \ldots \add E_n$ can be used to specify an 
expression $E_i$,  
together with the failure-handling routine $E_{i+1}$ for $i = 0,\ldots,n-1$.

It is well-known that traditional exception handling in functional languages 
of the conventional $try$-$catch$ style adds semantic complications to
languages \cite{Gov93}. For this reason, only a few functional languages such as Standard ML support 
exception handling.  As we will see, our scheme is immune to this problem.
Our construct is useful for the following reasons.

In sequential implementations, the single exception handler construct $A\add\ B$ can be used to handle single
exception. It can be seen as an optimized version
of traditional one with multiple exception handlers of the form $A\ handle\ B$. The need for this separate
construct is obvious:  although  $A \add\ B$ can be built from the 
existing $A\ handle\ B$, the resulting program are quite cumbersome, difficult to
 read/write/reason about. 

In parallel implementations, there is a more serious reason for the need of this construct.
 A key problem is that traditional mechanisms  lead to
semantical complications.  For example, suppose both $E_1$ and $E_2$ have single 
exception point,  namely $e_1$ and $e_2$ with exception handlers $h(e_1)$ and $h(e_2)$ respectively.
It is well-known then that
if we evaluate $E_1,E_2$ of an expression $f(E_1,E_2)\  handle\   h(e_1), h(e_2) $ in parallel,  it could  result in
nondeterministic behavior, depending on the order of evaluation of those subexpressions.
This violates the very property of functions!

This semantical problem is intrinsic with the concept and must be eliminated to
preserve clean and concise semantics of functional languages.
In ML, this problem is solved by $disallowing$ parallel evaluation of an expression.
However, this restriction is obviously unreasonable.  Worse,  considering that an error code is a global variable,
  the notion of inspecting (and possible subsequent updating) an exception is
similar to and as harmful as global variables in imperative languages.
 To overcome this problem, 
our approach is  to strictly restrict the number of handlers to $one$ for each function\footnote{Our approach
 allows many
kinds of exceptions though, so that proper error messages should be reported to the user.}.
This is to maintain the semantics as simple and clean as possible.
This restriction is not as severe as it looks, because  expressions with multiple
handlers can often be recursively transformed to expressions with single handler by
rearranging these handlers.
For example,   $f(E_1,E_2)\ handle\  h(e_1),h(e_2),h(e_3)$  can be rewritten as
 $f(E_1 \add  h(e_1), E_2 \add h(e_2)) \add h(e_3)$.  That is, if every programmer tries to
handle exceptions as locally as possible and to reduce the number of propagated errors,
our approach can approximately simulate existing exception handling mechanism.

\section{Sequential-conjunction  declarations}

SC declarations are intended to handle exceptions at the function declaration level.
  An
illustration of this  aspect is provided by the following SC definition of the
function $sort(X)$:

\begin{exmple}
$     (sort(X)$ ${\rm =}$ \> \hspace{3em} $heapsort(X)) \adc $    \\
 $ (sort(X)$ ${\rm =}$ \> \hspace{3em} $quicksort(X)) \adc $    \\
 $ (sort(X)$ ${\rm =}$ \> \hspace{3em}  $bubblesort(X)) $   \\
\end{exmple} 
\noindent
The  above definition  contains a SC declaration, denoted by $\adc$.
 As a particular example, evaluating $sort([3,100,40])$ would result in selecting and 
executing the first expression $heapsort([3,100,40])$.  If the heapsort module is available in the program, then the given goal 
will succeed, producing the solution. If the execution fails for some reason (type mismatch,
argument mismatch, heapsort not defined, etc), the machine tries the plan B, \ie, the quicksort module, and so on. 

SC declarations can also be used to define ordinary functions with a $case$. For example, the fibonacci function can be defined as:

\begin{exmple}
$     (fib(0)$ ${\rm =}$ \> \hspace{2em} $1)\ \adc $    \\
 $ (fib(1)$ ${\rm =}$ \> \hspace{2em} $1)\ \adc $    \\
 $ (fib(X)$ ${\rm =}$ \> \hspace{3em}  if $X > 1$ then $fib(X-1)+fib(X-1))$   \\
\end{exmple}

In addition, this idea of handling exceptions at the declaration level is closely related to
function overloading and polymorphic functions.
For example,  function overloading is a limited form of exception handling at the declaration level and
can be precisely captured by $\adc$.
In summary,  $\adc$ is a  construct which {\it unifies} -- with a clean semantics -- function overloading, a $case$ function and
exception handling.

\section{The Language}\label{s:logic}

The language is a version of the core functional languages --- also one of recursive functions ---
 with SC/SD operators. 
It is described
by $E$- and $D$-rules given by the abstract syntax as follows:
\begin{exmple}
\>$E ::=$ \>  $c \sep x \sep  h(E,\ldots,E) \sep     E \add E \sep \top \sep \bot(err)$ \\   \\
\>$D ::=$ \>  $f(t_1,\ldots,t_n) = E   \sep D \land D \sep D \adc D  $\\
\end{exmple}
\noindent
In the abstract syntax, $E$ and $D$ denote the expressions and the definitions, respectively.
In the rules above, $c$ is a constant, $x$ is a variable, $t$ is a term which is either a variable or a constant, and
$err$ is the error code. A set of function definitions $D$ with an  expression $E$ is called a program in this language.

\newcommand{\bc}{bc}

Following the traditional approach for defining semantics \cite{Mil89jlp,KK07}, we will present the semantics of this language, essentially an interpreter for the language, as a set of rules in Definition 1.
The evaluation strategy assumed by these rules is an eager evaluation. 
 Note that execution  alternates between 
two phases: the evaluation phase defined by \textit{eval}
and the backchaining phase by \textit{bc}. 

In  the evaluation phase, denoted by $\intp(D,E,K)$, the machine tries to evaluate an expression $E$ from the program $D$, a set of definitions, to get a value $K$. The rules (7) -- (10) 
are related to this phase. Note that these rules written in logic-programming style, \ie, $\intp(D,E,K)$ is true if the evaluation result of $E$ in $D$ is $K$. 
For instance, if $E$ is a function call $h$, the machine first evaluates all of its arguments and then looks for a definition of $h$ in the program in the backchaining mode (Rule 6).


The rules (1) -- (5) describe the backchaining mode, denoted by $bc(D_1,D,h,K)$.
In the backchaining mode, the machine tries 
to evaluate a function call $h$
by using the function definition in the program $D_1$.

\begin{defn}\label{def:semantics}
Let $E$ be an expression  and let $D$ be a program.
Then the notion of   evaluating $\lb D,E\rb$ to a value $K$ --- $\intp(D,E,K)$ --- 
 is defined as follows:

\begin{numberedlist}


\item    $\bc(h(c_1,\ldots,c_n) = E, D, h(c_1,\ldots,c_n), K)$ \\ if 
 $\intp(D, E, K)$. \% switch to evaluation mode. 

\item    $\bc(D_1\land D_2,D,h(c_1,\ldots,c_n),K)$  \\
 if   $\bc(D_1,D,h(c_1,\ldots,c_n),K)$. \% look for $h$ in $D_1$

\item    $\bc(D_1\land D_2,D,h(c_1,\ldots,c_n),K)$  \\
  if   $\bc(D_2,D,h(c_1,\ldots,c_n),K)$. \% look for $h$ in $D_2$

 \item    $\bc(D_1\adc D_2,D,h(c_1,\ldots,c_n),K)$  \\
   if   $\bc(D_i,D,h(c_1,\ldots,c_n),K)$, provided that $D_i(i=1\ \rm{or}\ 2)$ is the first successful declaration.
 \% SC declaration 

\item    $\bc(h(x_1,\ldots,x_n) = E, D, h(c_1,\ldots,c_n),K)$ \\
if   $\bc(h(c_1/x_1,\ldots,c_n/x_n) = E', D, h(c_1,\ldots,c_n),K)$ where
    $E' =   [c_1/x_1,\ldots,c_n/x_n]E$. \% argument passing to
 $h$ and $E$.

\item    $\intp(D,h(c_1,\ldots,c_n),K)$ \\ if   $\bc(D,D,h(c_1,\ldots,c_n),K )$. \% 
 switch to backchaining by making a copy of $D$ for a function call.

\item    $\intp(D,h(E_1,\ldots,E_n),K)$ \\ if $\intp(D,E_i,c_i)$ and $\intp(D,h(c_1,\ldots,c_n),K)$.
 \%  evaluate the arguments first.

\item    $\intp(D,\top,\top)$. \% 
 $\top$ is always a success.

\item   $\intp(D, c, c)$.  \% A success if $c$ is a constant.


\item $\intp(D,E_1\add E_2,K)$ \\ if  
$\intp(D,E_i,K)$, provided that $E_i(i = 1\ \rm{or}\ 2)$ is the first successful expression.
\%  exception handling. 

\end{numberedlist}
\end{defn}

\noindent  For simplicity, other popular constructs such as \textit{if-then-else} and pattern matching are not shown above.
 Note that evaluating an expression either returns a success with its value, or returns a
failure with an error message. If $\intp(D,E,K)$ has no derivation, it reurns a failure.
For example, $\intp(D,\bot(err),K)$ returns a failure, after  printing the error term to the user.

\section{Examples }\label{sec:modules}

As an  example, let us consider the well-known $\textit{div}$ function. 
There are many different ways to define this function and below is another one.

\begin{exmple}
$\textit{div}(x,y) =  (x/y) \add infinity$
\end{exmple}
\noindent  Evaluating $\textit{div}(4,2)$, the machine returns
a success with value 2.
On the other hand, evaluating $div(4,0)$ incurs backchaining. The machine tries $4/0$ first.
Since it leads to a failure, it tries a constant $infinity$. This leads to a success and 
the system returns a success with its value $\textit{infinity}$.

There are other ways to define the $\textit{div}$ function and below is another example.

\begin{exmple}
$\textit{div}(x,0) = \bot(divbyzero)\ \adc$ \\
$\textit{div}(x,y) = $ if $y > 0$ then $(x/y)$
\end{exmple}
\noindent  In the above, let us consider evaluating $\textit{div}(4,0)$. Using backchaining, it  tries 
to evaluate $\bot(divbyzero)$, which leads to a failure, with printing the ``divbyzero' to the user.

\section{Conclusion}\label{sec:conc}

In this paper, we proposed an extension to functional languages with  
SC/SD  operators. This extension allows expressions  $E_0 \add \ldots \add E_n$  where $E_0,\ldots,E_n$ are expressions, together with $D_0 \adc \ldots \adc D_n$  where $D_0,\ldots,D_n$ are declarations.
These expressions are 
particularly useful for specifying exception handling in a flexible way.

\section{Acknowledgements}

This work  was supported by Dong-A University Research Fund.

\bibliographystyle{ieicetr}


\end{document}